\documentstyle[preprint,aps,epsf,graphicx]{revtex}
\begin{document}

\tightenlines

\title{Partial wave contributions to the antikaon potential 
at finite momentum}

\author{Laura Tol\'os, Angels Ramos, Artur Polls}
\address{Departament d'Estructura i Constituents de la Mat\`eria,
Universitat de Barcelona, \\
Diagonal 647, 08028 Barcelona, Spain
}

\author{Thomas T.S. Kuo}

\address{
Department of Physics and Astronomy, State University of New York, \\
Stony Brook, NY 11794, USA}

\date{\today}

\maketitle
\begin{abstract}

The momentum dependence of the antikaon optical
potential in nuclear
matter is obtained from a microscopic and self-consistent 
calculation using the meson-exchange J\"ulich
$\bar{K}N$ interaction. 
Two self-consistent
schemes are discussed, which would lead to substantially different
predictions for the width of ${\bar K}$ nuclear bound states.
The effect of higher partial waves of the $\bar{K}N$ interaction,
beyond the $L=0$ component, is studied and
found to have moderate but non-negligible effects on the ${\bar
K}$ nuclear potential at zero momentum. At momenta as large as 500
MeV/c, relevant in the analysis of heavy-ion collisions, the
higher partial partial waves modify the ${\bar K}$ optical
potential by nearly a factor of two.

\end{abstract}
\vskip 0.5 cm



\section{Introduction}
\label{sec:intro}

Understanding the behavior of matter under extreme conditions of
density and temperature, such as those found in neutron stars or
in heavy-ion collisions, requires
a good knowledge and control on the 
modification of the properties of hadrons in the nuclear medium.
In particular, the in-medium effects on antikaons have received
much attention over the last years, especially after the speculation
of the possible existence of an antikaon condensed phase 
\cite{KN86}.

Empirical analysis of kaonic atoms \cite{FGB94} predict a strong
attractive $K^-$-nucleus potential, 
but, recently, the data have also been shown to be compatible
with a potential which is four times
less attractive \cite{satoru00}. This last potential was
obtained from a microscopic
self-consistent
calculation \cite{ramos00} using the chiral non-perturbative
model of ref.\cite{oset98} that reproduces the ${\bar K}N$
scattering observables. A recent fine-tuning of this potential 
\cite{baca}, was shown to produce a $\chi^2$ as low as that of
ref.~\cite{FGB94}. These observations imply that the data on kaonic atoms,
although giving a strong indication of
attraction,
do not really constrain the size of the $K^-$ optical potential at
nuclear matter
saturation density.

The dynamics of the $\bar{K}N$ interaction is particularly rich  due
to the presence of an isospin 0 resonance, the $\Lambda(1405)$, which
lies only 27 MeV below
the ${\bar K}N$ threshold. As a result, the isospin
averaged ${\bar K}N$ interaction in free space is in fact
repulsive.
It was soon pointed out, however, that in the medium the Pauli
blocking on nucleon states moves
the resonance to higher energies and makes the interaction attractive
\cite{Koch94,WKW96,Waas97}. This
particular (resonant-like) 
energy dependence of the ${\bar K}N$ interaction at subthreshold energies
requires an especially careful treatment of all in-medium
effects. In fact,
when the attraction felt by the antikaon is
self-consistently incorporated in the calculation \cite{Lutz98} it
compensates partly the effects induced by Pauli blocking and
the $\Lambda$ resonance gets broader and moves back to lower energies,
around its free space location.
Moreover, since
an extremely
important ingredient to generate the $\Lambda$ resonance is the
coupling of the ${\bar K}N$ system to the $\pi\Sigma$ one, 
the modification of the pion properties was also
included in ref.~\cite{ramos00},
with the result that the resonance width increased further more and
the peak shifted upwards to an
energy slightly above the free space one.

Heavy-ion collisions can also provide information on the modification
of hadronic properties. In particular the enhanced $K^-/K^+$ ratio
measured at GSI \cite{kaos97} can also be understood assuming that the
antikaons feel a strong attraction \cite{cassing97,Li97,Li98,sibirt98}.
This
lowers the
threshold for $K^-$ production, hence enhancing the ratio at
subthreshold energies. In a recent work \cite{schaffner00}, it has been
shown that the
increase on this ratio could also be explained from the in-medium
enhanced production
of antikaons through $\Sigma$ hyperons due to the shifting of the
$\Lambda(1405)$ resonance to higher energies, a mechanism
that was already suggested in ref.~\cite{Ohnishi97} to explain 
the data of $K^-$ absorption at rest on $^{12}$C
\cite{Tamura89,Kubota96}.

The situation on the antikaon properties in the medium is, at present, far
from being clear. Although it is
commonly accepted that to explain kaonic atom data and
the results on heavy-ion collisions the  $K^-$ must feel
some attraction, the magnitude of it has not been really determined by
the available observations.

One may aim at extracting this information from theoretical models that
have been well constrained in free space by the ${\bar K}N$
scattering observables. Due
to the special dynamics of the ${\bar K}N$ interaction, these models must
treat carefully all medium effects,
as well as the energy and momentum dependences of the in-medium ${\bar
K}N$
interaction. The momentum dependence of the ${\bar K}$ optical potential 
is  particularly
important in the analysis of heavy-ion collisions since
antikaons are created at a finite momentum of around $250-500$ MeV/c.

Intimately connected with the momentum
dependence is the role of higher partial waves of the
$\bar{K}N$ interaction. For a relative momentum of the
${\bar K}N$ pair as large as that explored in heavy-ion collisions,
the importance of
the higher partial waves, beyond the commonly considered 
$L=0$ component, might become quite relevant.

The purpose of the present work is to
study all these effects using the ${\bar K}N$ interaction
of the J\"ulich group \cite{holinde90}, which is based on meson-exchange.
The J\"ulich potential predicts the $\Lambda(1405)$
resonance to be
a quasibound 
${\bar K}N$ state and reproduces satisfactorily the ${\bar K}N$
scattering
observables. The matrix elements of this interaction depend
on the incident and outgoing three-momenta, as well as
on the total invariant center-of-mass energy. A partial
wave decomposition is also 
given in ref.~\cite{holinde90}, hence allowing the investigation 
on how the different partial waves contribute to
the ${\bar K}$ optical potential in the nuclear medium. 

The paper is organized as follows. In Sect. \ref{sec:medium} we present
the formalism to derive the in-medium effective $\bar{K}N$ interaction and
the corresponding ${\bar K}$ optical potential. Our results are
shown in Sect. \ref{sec:results}, where particular attention is paid 
to study the effects of the higher partial waves on the ${\bar
K}$ optical potential as obtained with two different
self-consistent schemes.
Our conclusions are summarized in Sect. \ref{sec:conclu}.

\section{In-medium ${\bar K}N$ interaction and ${\bar K}$ optical
potential in nuclear matter}
\label{sec:medium}

In this section we present the formalism to obtain
the self-energy or single-particle potential
 of a $\bar{K}$ meson embedded in infinite symmetric nuclear matter.
This self-energy accounts for the interaction of the $\bar{K}$
with the nucleons and
its calculation requires the knowledge of the in medium $\bar{K}N$
interaction, which we will describe by a $G$-matrix.
The medium effects incorporated in this $G$-matrix include
the Pauli blocking on the nucleons in the intermediate states as well as 
the
dressing of the $\bar{K}$ and the different baryon lines. We will explore
different
ways of dressing the intermediate states in solving the $G$-matrix, which
will define the type  of self-consistency of the calculation
giving rise to the different approaches
discussed in this paper.

\subsection{Effective $\bar {K} N$ interaction}

The effective $\bar{K} N$ interaction in the nuclear medium or
$G$-matrix is obtained from the bare $\bar {K} N$ interaction derived
in the meson exchange framework \cite{holinde90}. As the bare
 interaction allows for the transition from the $\bar {K} N$
channel to the $\pi \Sigma$ and $\pi \Lambda$ channels,
all having strangeness $S=-1$, we are confronted with a coupled channel
problem.
The resultant meson-baryon (MB) $G$-matrices can be grouped in a matrix
notation, where each box corresponds to one channel.
The $\bar {K} N$ channel can have isospin $I=0$ or $I=1$. In the first
case, it can only couple  to the $\pi \Sigma$ channel and the
corresponding
matrix has the following structure

\[\left(\begin{array}{cc}
G_{{\bar{K}}N\rightarrow{\bar{K}}N} &
G_{{\bar{K}}N\rightarrow{\pi}{\Sigma}} \\
G_{{\pi}{\Sigma}\rightarrow{\bar{K}}N} &
G_{{\pi}{\Sigma}\rightarrow{\pi}{\Sigma}}
        \end{array}
\right) \ ,\]
while for $I=1$ it can couple to both the $\pi \Sigma$ and $\pi \Lambda$ 
channels

\[\left(\begin{array}{ccc}
G_{{\bar{K}}N\rightarrow{\bar{K}}N} &
G_{{\bar{K}}N\rightarrow{\pi}{\Sigma}} &
G_{{\bar{K}}N\rightarrow{\pi}{\Lambda}}  \\
G_{{\pi}{\Sigma}\rightarrow{\bar{K}}N} &
G_{{\pi}{\Sigma}\rightarrow{\pi}{\Sigma}} &
G_{{\pi}{\Sigma}\rightarrow{\pi}{\Lambda}}  \\
G_{{\pi}{\Lambda}\rightarrow{\bar{K}}N} &
G_{{\pi}{\Lambda}\rightarrow{\pi}{\Sigma}} &
G_{{\pi}{\Lambda}\rightarrow{\pi}{\Lambda}}
        \end{array}
  \right) \ .\]

Keeping this structure in mind, the $G$-matrix is formally given by

\begin{eqnarray}
\langle M_1 B_1 \mid G(\Omega) \mid M_2 B_2 \rangle &&= \langle M_1 B_1
\mid V({\sqrt s}) \mid M_2 B_2 \rangle   \nonumber \\
&& \hspace*{-2cm}+\sum_{M_3 B_3} \langle M_1 B_1 \mid V({\sqrt s }) \mid
M_3 B_3 \rangle
\frac {Q_{M_3 B_3}}{\Omega-E_{M_3} -E_{B_3}+i\eta} \langle M_3 B_3 \mid
G(\Omega)
\mid M_2 B_2 \rangle
   \label{eq:gmat1}
\end{eqnarray}
In Eq. (\ref{eq:gmat1}),  $M_i$ and $B_i$  represent the possible
mesons ($\bar {K}$, $\pi$) and
baryons ($N$, $\Lambda$, $\Sigma$), respectively, and their corresponding
quantum numbers such as spin, isospin, strangeness, and linear momentum.
The function $Q_{M_3,B_3}$ stands for the Pauli operator which allows
only intermediate nucleon states
compatible with the Pauli principle. The energy variable $\Omega$ is
the
so called starting energy that we will always take as a real
quantity, while $\sqrt{s}$ is the invariant
center-of-mass energy which enters into  the definition of the bare
potential \cite{holinde90}, i.e.
$\sqrt{s}=\sqrt{\Omega^2-P^2}$, where $(\Omega,\vec{P}\,)$ is
the total meson-baryon fourmomentum in a frame in which nuclear matter is at
rest.

The former equation for the $G$-matrix has to be considered together with
a prescription for the
single-particle energies of all the mesons and baryons participating
in the reaction and in the intermediate states. These energies can be
written as
\begin{equation}
 E_{M_i(B_i)}(k)=\sqrt{k^2 +m_{M_i(B_i)}^2} + U_{M_i(B_i)}
(k,E_{M_i(B_i)}^{qp}) \ ,
\label{eq:spen}
\end{equation}
where $U_{M_i(B_i)}$ is the  single-particle
potential of each meson (baryon) calculated at the real
quasi-particle energy $E_{M_i(B_i)}^{qp}$, obtained by solving  the
following
equation
\begin{equation}
E_{M_i(B_i)}^{qp}(k)=\sqrt{k^2 +m_{M_i(B_i)}^2} + {\mathrm {Re}\,} 
U_{M_i(B_i)}
(k,E_{M_i(B_i)}^{qp}) \ .
\label{eq:qp}
\end{equation}

 In the present paper we have
considered the single-particle potential for the $\bar K$ meson
and all baryons. While for the $\Lambda$ and $\Sigma$ hyperons we use the
simplified form
\begin{equation}
U_{\Lambda,\Sigma}=-30\frac{\rho}{\rho_{0}},
\end{equation}
where $\rho_{0}$ is the saturation nuclear matter density, for nucleons 
we take the density dependent parametrization \cite{wirin88}
\begin{equation}
U_{N}(\rho,k)
=\alpha(\rho)+\frac{\beta(\rho)}{1+\left[
\displaystyle\frac{k}{\gamma(\rho)}\right]^2}
\ ,
\label{eq:param}
\end{equation}
that was
fitted to the single-particle potentials obtained from variational
calculations using realistic $NN$ interactions. The density-dependent
real parameters $\alpha(\rho)$, $\beta(\rho)$ and $\gamma(\rho)$ are
given in Table I of ref. \cite{wirin88}. The nucleon potential at 
$\rho=\rho_0$ amounts to $-70$ MeV.

In the Brueckner-Hartree-Fock approach, the $\bar K$ self-energy  is
schematically given by
\begin{equation}
 U_{\bar K}(k,E_{\bar K}^{qp})= \sum_{N \leq F} \langle \bar K N \mid
 G_{\bar K N\rightarrow
\bar K N} (\Omega = E^{qp}_N+E^{qp}_{\bar K}) \mid \bar K N \rangle,
\label{eq:self}
\end{equation}
where the summation over nucleon states is limited by the nucleon Fermi
momentum.

At the required ${\bar K}$ energies, the
$G$-matrix in the above equation becomes complex
due to the possibility of $\bar{K}$$N$ decaying into the $\pi\Sigma$,
$\pi\Lambda$ channels.
As a consequence, the potential
$U_{\bar K}$ is also a complex quantity. The explicit calculation
of $U_{\bar K}$ will be discussed in the next subsection.

The influence
 of the medium on the $G$-matrix is studied by considering different
approaches
to  the self-consistent single-particle energy used for the $\bar{K}$
meson in Eq.~(\ref{eq:gmat1}). We have
taken two prescriptions:
a) the real quasi-particle energy defined in Eq. (\ref{eq:qp}) and
b) the complex $E_{\bar K}$ defined in Eq. (\ref{eq:spen}) which
also contains the
 imaginary part of $U_{\bar K}$ calculated at $E_{\bar K}^{qp}$.

We solve the $G$-matrix equation by doing first a partial wave
decomposition. Using the quantum numbers of the
relative and
center-of-mass motion, the $G$-matrix equation in this basis reads
\begin{eqnarray}
\langle (M_1B_1);k''|
      G^{L J I}(P,\Omega) | (M_2B_2);k  \rangle
     &&=  
     \langle (M_1B_1);k'' |
      V^{L J I}({\sqrt{s}}) | (M_2B_2);k \rangle 
      \nonumber \\
    &&\hspace*{-5cm} +\sum_{M_3 B_3}
      \int k'^{2}dk'
      \langle (M_1 B_1);k'' |
      V^{L J I}({\sqrt{s}}) | (M_3 B_3); k'\rangle
\nonumber
\\
      && \hspace*{-4.5cm}\times \frac{\overline{Q}_{M_3 B_3}(k',P)}{\Omega
-\sqrt{M_{B_{3}}^2+\widetilde{k_{B_{3}}^2}}                             
-
\sqrt{M_{M_{3}}^2+\widetilde{k_{M_{3}}^2}}
-U_{B_{3}}(\widetilde{k_{B_{3}}^2})-U_{M_{3}}(\widetilde{k_{M_{3}}^2})
+i\eta}
\nonumber \\
      && \hspace{-4.5cm}\times \langle (M_3 B_3);k' |
      G^{L J I}(P,\Omega) | (M_2B_2);k \rangle \ ,
   \label{eq:gmat}
\end{eqnarray}
where the variables $k$, $k'$, $k''$ and $L$  denote
relative linear momenta
and orbital momentum, respectively, while $P$ is the linear
center-of-mass momentum.
The functions $\widetilde{k_{B}^2}$ and $\widetilde{k_{M}^2}$ are,
respectively, the square of the momentum of the baryon and that of the
meson in the intermediate states, averaged
over the angle between the total momentum ${\vec P}$ and the relative
momentum $\vec{k}'$ (see the definitions in appendix A)
\begin{eqnarray}
\widetilde{k_B^2}(k',P)&=&k'^2 + \left(\frac{M_B}{M_{
M}+M_B}\right)^2
P^2 \\
\widetilde{k_{M}^2}(k',P)&=&k'^2 + \left(\frac{M_{M}}{M_{
M}+M_B}\right)^2  
P^2 \ .
\end{eqnarray}
The angle average of the Pauli operator, $\overline{Q}_{M_3 B_3}(k',P)$,
is shown explicitly in appendix A and differs
from unity only in the case of the ${\bar K}N$ channel.
The total angular momentum and isospin
are denoted by $J$ and $I$,
respectively.
For each $J$, two values of orbital angular momentum, $L=J+1/2$ and
$L=J-1/2$,  are allowed. However, due to parity conservation, the
interaction can not mix these two states and,
as a consequence, the orbital angular momentum is also a conserved
quantity.

\subsection{$\bar{K}$ single-particle energy in
Brueckner-Hartree-Fock approximation}

By using the partial wave decomposition of the $G$-matrix, the Brueckner-Hartree-Fock approximation to the single-particle potential
of a $\bar K$ meson embedded in a Fermi sea of nucleons 
[Eq.~(\ref{eq:self})] becomes

\begin{eqnarray}
 U_{\bar{K}}(k_{\bar{K}},\omega=E_{\bar K}^{qp}(k_{\bar K})) &=
& \frac{1}{2}
\sum_{L, J, I}(2J+1)(2I+1) (1+\xi)^3 \int_{0}^{k_{max}}k^2dk
f(k,k_{\bar{K}}) \nonumber \\
& & \times  \langle (\bar{K}N) ; k | G^{L J I}
(\overline{P^2},E^{qp}_{\bar{K}}(k_{\bar{K}})+ 
E^{qp}_{N}(\overline{k_{N}^2}))
|
(\bar{K} N); k \rangle  \ ,
\label{eq:upot1}
\end{eqnarray}
where $\overline{P^2}$ and $\overline{k_{N}^2}$ are the square 
of the center-of-mass momentum and nucleon momentum,
respectively, averaged
over the angle between the external ${\bar K}$ momentum in the lab
system, $\vec{k}_{\bar K}$, and the ${\bar K}N$
relative momentum, ${\vec k}$, used as integration variable in Eq.
(\ref{eq:upot1}) (see appendix B for details). These angle averages
eliminate the angular dependence
of the $G$-matrix and allow to perform the angular integration in Eq.
(\ref{eq:upot1}) analitically, giving rise to the weight function,
$f(k,k_{\bar{K}})$,
\begin{equation}
f(k,k_{\bar{K}})= \left\{ \begin{array}{cl} 1 & {\rm for\ }  k\leq
\frac{k_{F}-\xi k_{\bar{K}}}{1+\xi} , \\ 0 &
{\rm for\ }
|\xi k_{\bar{K}}-(1+\xi)k| > k_{F} , \\
\displaystyle\frac{{k_{F}}^2-[\xi k_{\bar{K}}-(1+\xi)k]^2}{4\xi
(1+\xi)k_{\bar{K}}k
} & \mbox{otherwise,}
\end{array} \right. \nonumber
\end{equation}
where $\xi=\displaystyle\frac{M_{N}}{M_{\bar{K}}}$ and $k_F$ is the
Fermi momentum.
The magnitude of
the relative momentum 
$k$ is constrained by
\begin{equation}
   k_{max} = \frac{k_{F}+\xi k_{\bar{K}}}{1+\xi} \ .
\nonumber
\end{equation}

After self-consistency for the on-shell value
$U_{\bar K}(k_{\bar K},E_{\bar K}^{qp})$ is
achieved, one can obtain the complete energy dependence of the
self-energy,
$U_{\bar K}(k_{\bar K},\omega)$, which can be used to determine the
 $\bar K$ single-particle propagator in the medium,
\begin{equation}
D_{\bar K}(k_{\bar K},\omega) = \frac {1}{\omega^2 -k_{\bar K}^2 -m_{\bar
K}^2
-2 m_{\bar K} U_{\bar K}(k_{\bar K},\omega)} \ ,
\label{eq:prop}
\end{equation}
and the corresponding spectral density
\begin{equation}
S_{\bar K}(k_{\bar K},\omega) = - \frac {1}{\pi} {\mathrm Im\,} D_{\bar
K}(k_{\bar K},\omega) .
\label{eq:spec}
\end{equation}

\section{Results}
\label{sec:results}

\begin{figure}[htb]
\centerline{
     \includegraphics[width=0.6\textwidth]{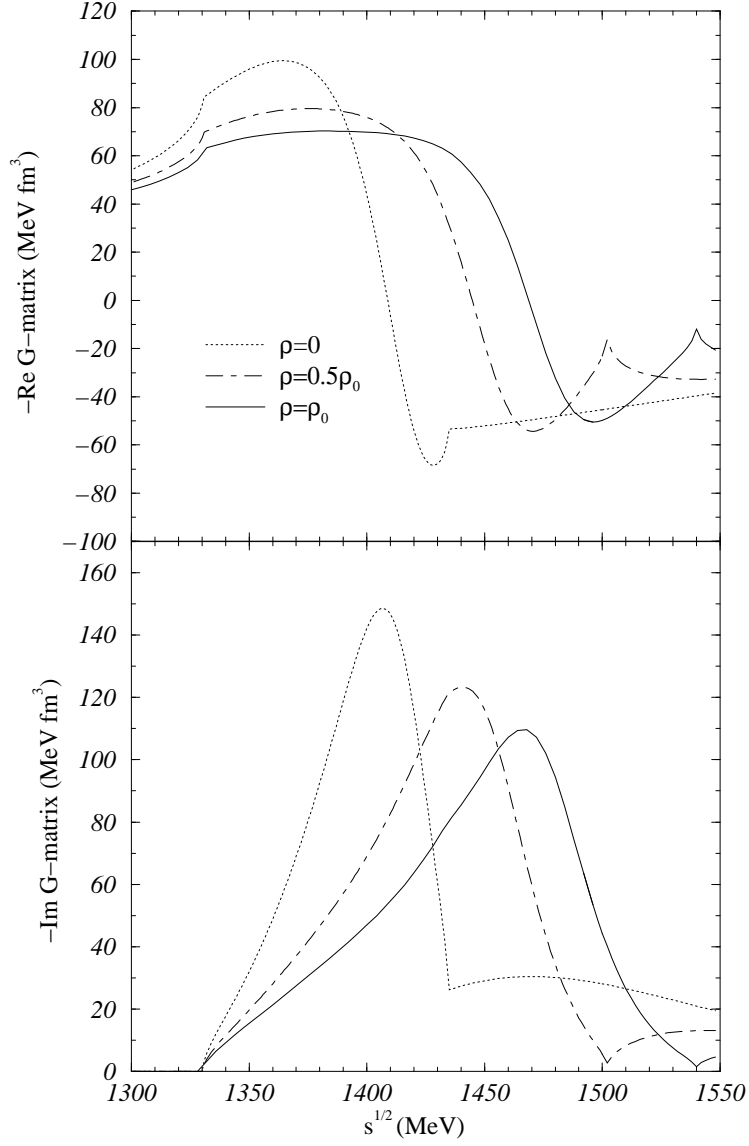}
}
      \caption{\small 
Real and imaginary parts of the ${\bar K}N$ amplitude in the
$I=0$, $L=0$ channel as functions of the center-of-mass energy at total 
momentum $|\vec k_{\bar K} +\vec k_{N}|=0$ for various
densities.}
        \label{fig:kaon1}
\end{figure}

We start this section by showing explicitly the most
characteristic medium modification of the ${\bar K}N$ amplitude, which
is obtained when the Pauli blocking on the intermediate nucleons
is incorporated. The real and imaginary parts of the
resulting in-medium ${\bar K}N$ amplitudes 
for a total momentum $\mid \vec{k}_{\bar K} + \vec{k}_N \mid=\vec{0}$ 
are shown
in Fig.~\ref{fig:kaon1} 
as functions of the invariant center-of-mass energy, for three different
densities:
$\rho=0$ (dotted lines), $\rho=\rho_0/2$ (dot-dashed lines),
and $\rho=\rho_0$ (solid lines),  with $\rho_0=0.17$ fm$^{-3}$ being
the saturation density 
of nuclear matter.
We clearly see, as
noticed already by all earlier microscopic
calculations, the repulsive effect on the resonance produced by
having moved the threshold of intermediate allowed ${\bar K}N$ states
to higher energies, as a result of Pauli
blocking acting on the nucleon.

Clearly, this shift of the resonance changes the
${\bar K}N$ interaction at threshold ($\sim 1432$ MeV) 
from being repulsive
in free space to being attractive in the medium.
Since this effect is intimately connected with the
strong energy
dependence of the ${\bar K}N$ interaction,
important changes can be expected from a  
self-consistent incorporation of the ${\bar K}$ 
properties on the ${\bar K}N$ $G$-matrix, as
already noted by Lutz \cite{Lutz98} and confirmed in 
ref.~\cite{ramos00}.

This type of self-consistent calculations are also common in
Brueckner-Hartree-Fock studies of the $NN$ interaction,
where it is costumary to take the nucleon single-particle energies
as real quantities. This amounts to disregard the imaginary part of
the in medium $NN$ amplitude in the calculation of the
single-particle energy for nucleons above the Fermi momentum.
We have attempted a similar type of approximate scheme
for the ${\bar K}$ meson, although in this case the ${\bar K}N$ amplitude
is already complex at threshold due to the
$\pi\Lambda$, $\pi\Sigma$ channels. In this simplified
self-consistent scheme, only the real part of the
${\bar K}$ potential is retained in the solution of the $G$-matrix, i.e.
the ${\bar K}$ energy appearing in Eq. (\ref{eq:gmat1}) is taken to be
the real quasi-particle energy of Eq. (\ref{eq:qp}).

\begin{figure}[htb]
\centerline{
     \includegraphics[width=0.6\textwidth]{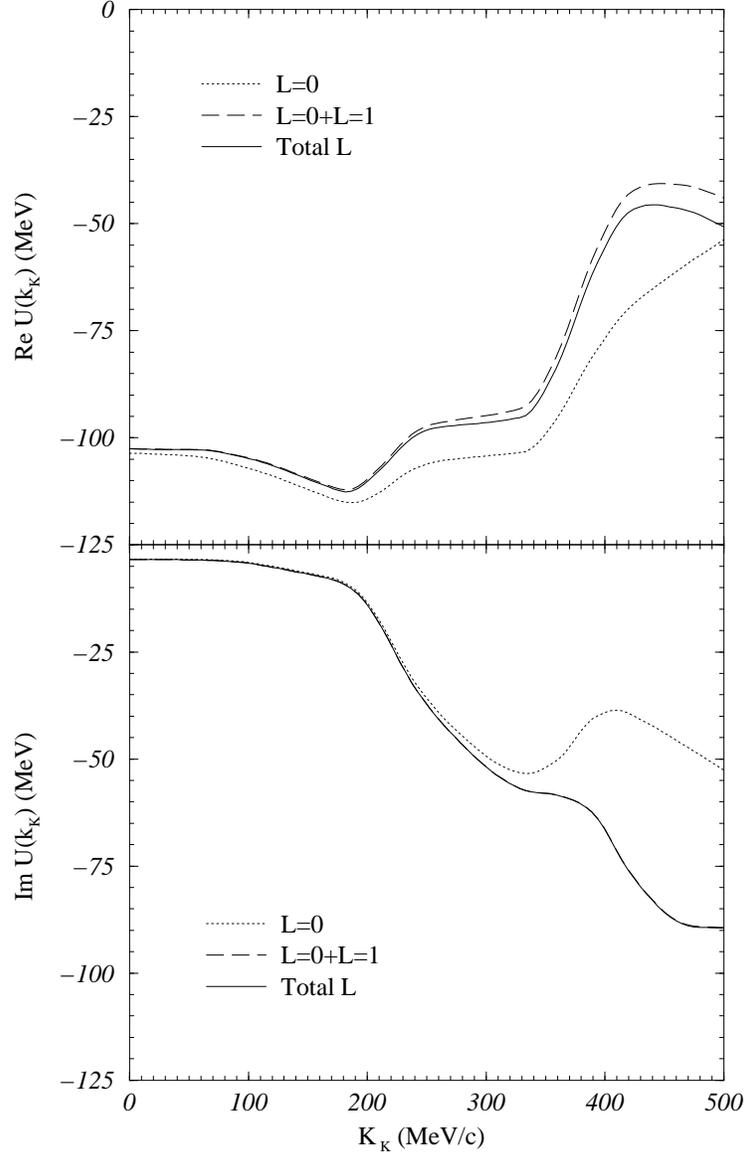}
}
      \caption{\small 
Real and imaginary parts of the ${\bar K}$ optical potential
at $\rho=\rho_0$ as functions of the antikaon momentum.
These results are obtained when only the real part of the
${\bar K}$ potential is determined self-consistently.}
        \label{fig:kaon2}
\end{figure}

Once self-consistency is achieved, we calculate both real and
imaginary
parts of the ${\bar K}$ potential, which are displayed
in Fig.~\ref{fig:kaon2} as
functions of the antikaon momentum, for
normal nuclear
matter density $\rho=\rho_0$. 
We explicitly show the
separate contribution of the various partial waves of the ${\bar K}N$
interaction. At zero momentum the real part of the ${\bar K}$ potential is
about $-100$ MeV and the partial waves higher than the $L=0$ component
give a negligible contribution. Both the real and imaginary parts of 
the potential
show some structure between $200$ MeV/c and $400$ 
MeV/c. This is connected to the behavior of the in-medium ${\bar K}N$ 
amplitude
which, in this simplified self-consistent scheme, still retains 
the resonant-like shape at $\rho=\rho_0$. 
We observe that the high partial waves
start being non-negligible
for momenta larger than $200$ MeV/c.
It is interesting to focus our attention on the imaginary part of the
${\bar K}$ potential which at low momentum is rather small. This is due
to the supressed coupling to intermediate $\pi\Sigma$ states as a result of
the strong attraction felt by the antikaon combined with the
attraction of around $-70$ MeV felt by the nucleon.
The in-medium ${\bar K}N$ system then explores energies below 
those available for $\pi\Sigma$ excitations and the antikaon width
becomes extremely small. In this low momentum region the
imaginary part of the ${\bar K}$ potential is essentially due to the
coupling to $\pi\Lambda$ states in the relatively weak $I=1$ channel. As
seen in the figure, for the $\pi\Sigma$ channel to start giving a
contribution to the imaginary part,
it is necessary
to provide the ${\bar K}$ with a finite momentum (around 200 MeV/c), such
that the condition
$E^{qp}_{\bar K} + E^{qp}_N > m_\pi + m_\Sigma + U_\Sigma \simeq 1300$ MeV
can be fulfilled.

The substantial attraction obtained for the $\bar K$ optical potential
combined with the
extremely small imaginary component of the ${\bar K}$ optical
potential might induce to think that
the chances of producing very narrow, hence observable, deeply bound
${\bar K}$-nucleus states are quite high. If these results were
confirmed by more sofisticated treatments of the in-medium effects,
an experimental search for these states should be indeed encouraged.
However, as we will show in the following, the use of a more realistic
self-consistent scheme wears this finding off.

Our second method consists of determining the complete complex ${\bar
K}$ single-particle energy given in Eq. (\ref{eq:spen})
self-consistently.
That means that the dressing of the ${\bar K}$ in the
intermediate states of the $G$-matrix equation is taken into account 
through a complex but energy-independent self-energy. This gives rise
to the
so-called ``quasi-particle''
spectral density, which has the energy dependence of 
a Lorentzian distribution.
Although this is an approximation with respect to the more sophisticate
attempts made in Refs. \cite{ramos00,Lutz98}, where the full
energy dependence
of the ${\bar K}$ self-energy is self-consistently determined, it is
sufficiently good for the studies we are carrying out in the present
work.
Moreover, at the end of the self-consistent scheme, we 
evaluate the spectral density of the ${\bar K}$ keeping the complete
momentum and energy dependence of the ${\bar K}$ self-energy.

\begin{figure}[htb]
\centerline{
     \includegraphics[width=0.6\textwidth]{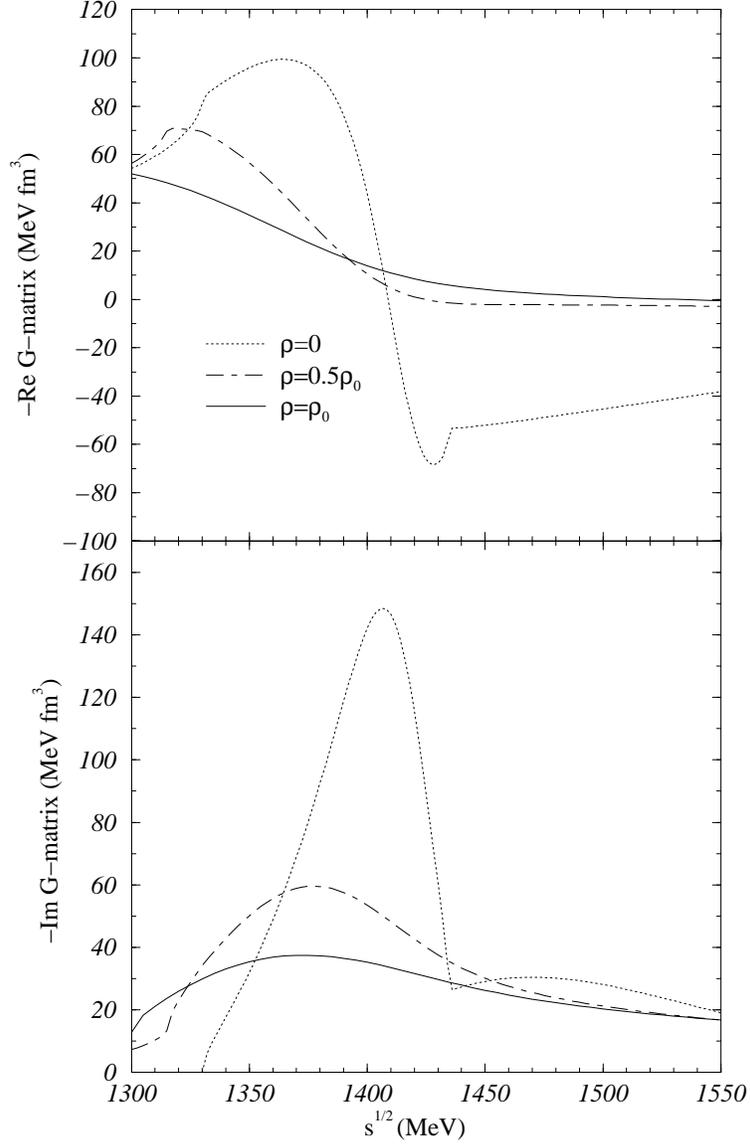}
}
      \caption{\small 
Real and imaginary parts of the ${\bar K}N$ amplitude in the
$I=0$, $L=0$ channel as functions of the center-of-mass energy at total 
momentum $|\vec k_{\bar K} +\vec k_{N}|=0$ for various
densities, as obtained from a self-consistent calculation with a complex 
${\bar K}$ optical potential.}
        \label{fig:kaon3}
\end{figure}

We first show, in Fig.~\ref{fig:kaon3}, the effect of dressing the
antikaon with a complex
self-energy on
the in medium $\bar{K}N$ amplitude. Comparing with the
results shown in Fig.~\ref{fig:kaon1}, we observe that the resonance peak
stays now pretty close or even lower than its free space location. As
mentioned above, this is due to the attraction felt by the ${\bar K}$
that compensates the repulsive effect induced by Pauli blocking on the
nucleon. Secondly, both real and imaginary parts of the ${\bar K}N$
amplitude become much smoother, as a result of the ${\bar K}$ strength
being spread out over energies. The resonance gets wider and dilutes much
earlier with increasing density. We note that these effects are in
total agreement with those obtained with the
self-consistent approaches of refs.~\cite{ramos00,Lutz98}.

\begin{figure}[htb]
\centerline{
     \includegraphics[width=0.6\textwidth]{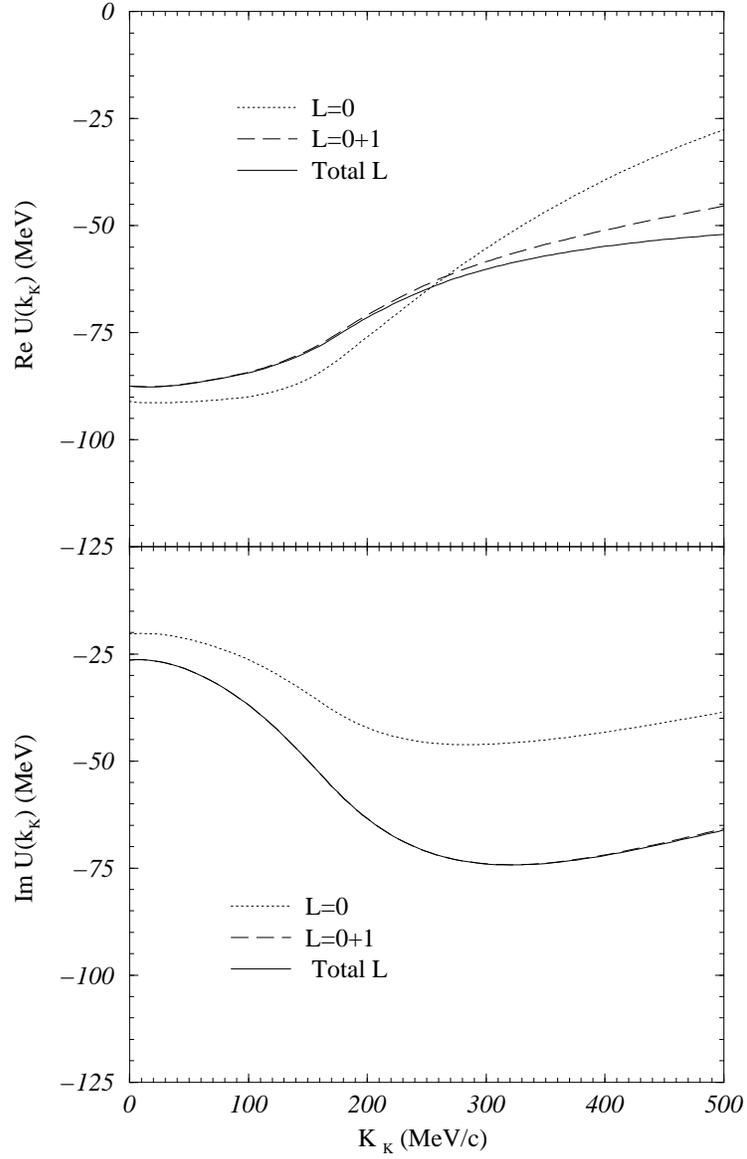}
}
      \caption{\small
Same as Fig.~\ref{fig:kaon2} but obtained from a
self-consistent scheme which uses the complex ${\bar K}$ optical
potential.}
        \label{fig:kaon4}
\end{figure}

Our results for the antikaon potential at $\rho=\rho_0$ obtained with this
complete self-consistent scheme are shown in
Fig.~\ref{fig:kaon4}. 
The real part of the ${\bar K}$ potential at zero momentum
increases to  $-87$ MeV.
It is worth noticing that the new scheme has 
produced a drastically different imaginary part.
The small value of
$-3$ MeV at zero momentum obtained using the
simplified  scheme turns out to be now around $-25$ MeV, hence making the
observation of bound ${\bar K}$ nuclear states more difficult. 
This is a consequence of the fact that the
in-medium ${\bar K}N$ amplitude becomes smoother and wider
when the ${\bar K}$ energies are complex. Since the ${\bar K}$
spectral density develops a width and, in turn, the ${\bar K}$ feels 
a reduced attraction, the ${\bar K}N$ states can
couple more easily to the $\pi\Sigma$ states than in the previous
self-consistent scheme. From these results one must conclude 
that any approach claiming for narrow
bound ${\bar K}$ nuclear states must be looked at with caution, 
because the self-consistent scheme affects enormously
the predicted ${\bar K}$
properties in the medium.

The effect of including the higher partial waves of the ${\bar K}N$
interaction is also seen in Fig.~\ref{fig:kaon4}. We observe that at zero
${\bar K}$ momentum there is already some contribution of partial waves
higher than $L=0$ due to the fact that the ${\bar K}$ meson interacts
with nucleons that occupy states up
to the Fermi momentum, giving rise to finite ${\bar K}N$
relative momenta of up to around 90 MeV/c. Clearly, the effect of the
higher partial waves increases with increasing ${\bar K}$ momentum,
flattening out the real part of the optical potential and producing more
structure to the imaginary part. At a ${\bar K}$ momentum around 500 MeV/c,
the inclusion of higher partial waves makes the real part more attractive,
from
$-28$ MeV to $-52$ MeV, and practically doubles the size of the imaginary
part.

\begin{figure}[htb]
\centerline{
     \includegraphics[width=0.6\textwidth]{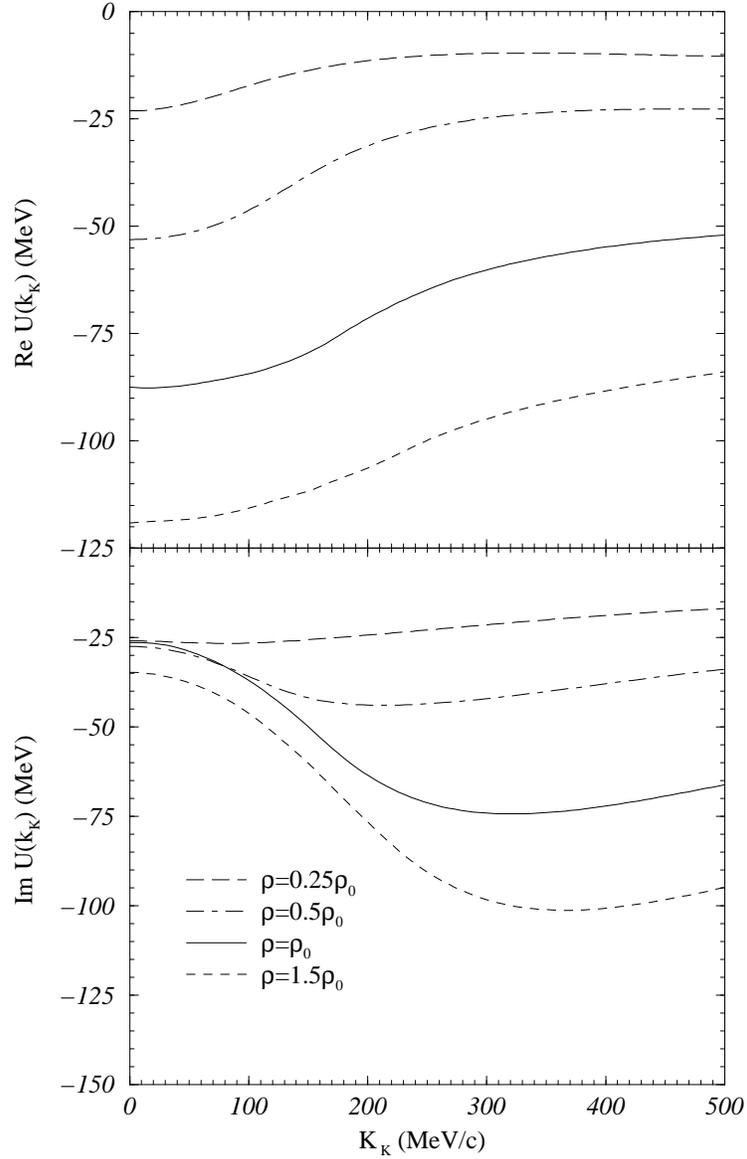}
}
      \caption{\small
Real and imaginary parts of the ${\bar K}$ optical potential
as functions of the antikaon momentum for various densities.}
        \label{fig:kaon5}
\end{figure}

In Fig.~\ref{fig:kaon5} we represent the real and the imaginary part of
$U_{\bar{K}}$ for different
densities, obtained with the scheme that considers the complex potential 
for the antikaon. As we
increase the density we obtain a more attractive potential due to two
facts. On one hand, as discussed previously, the interaction is more
attractive at higher densities. On the other hand, the $\bar{K}$ feels the
interaction of many more nucleons.
The momentum dependence is moderate for the real part. In contrast, the
imaginary
part for densities around the saturation density and beyond shows
more structure. This momentum dependence is smoother than that
obtained in ref.~\cite{sibirt98} from a phenomenological model using the
information of the vacuum ${\bar K}N$ amplitudes, where
the antikaon optical potential at normal nuclear matter density $\rho_0$
increases from $-140$ MeV at zero momentum to around $-50$ MeV at high
momenta. 

Starting from the self-consistent ${\bar K}$ optical potential shown in
Fig.~\ref{fig:kaon3} and reproduced by the solid line in
Fig.~\ref{fig:kaon5}, we perform two different tests in order to check
some of the approximations used in the literature. First, we compute the
optical potential neglecting the nucleon recoil corrections, as done in
ref.~\cite{schaffner00}. This amounts to calculate the ${\bar K}$
potential at momentum $k_{\bar K}$ from a $G\rho$ type expression, with
$G$ being the in-medium ${\bar K}N$
interaction between a ${\bar K}$ of momentum $k_{\bar K}$ and a nucleon
of zero momentum. The results are displayed by the dashed lines in
Fig.~\ref{fig:kaon5}. While the real part is barely affected by this
simplified approximation, the imaginary part shows some non-negligible
differences at low momenta. Secondly, we have neglected, in addition, the
nucleon
single-particle potential, which has been claimed to be unimportant in
ref.~\cite{schaffner00}. Note that in this later reference a small
imaginary part of around 10 MeV for the nucleon optical potential is
used, but the real
part is set to zero. The results are shown by the dotted line. The
differences with the dashed line
are quite appreciable at low momenta,
both for the real and imaginary parts of the ${\bar K}$ optical
potential. 

\begin{figure}[htb]
\centerline{
     \includegraphics[width=0.6\textwidth,angle=-90]{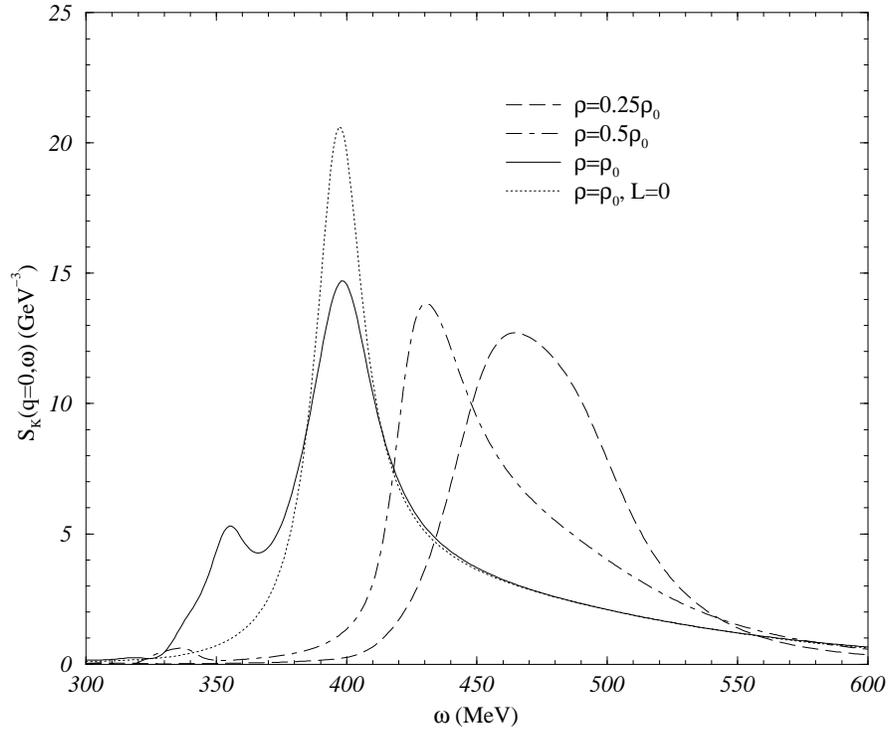}
}
      \caption{\small
$\bar{K}$ spectral density at $k_{\bar{K}}=0$ as a function of energy
 for various densities.}
        \label{fig:kaon7}
\end{figure}

Finally, once self-consistence is reached, we calculate
the full
energy dependence of the $\bar K$ self-energy which defines the in-medium
$\bar{K}$ single-particle propagator and its spectral density through
Eqs.~(\ref{eq:prop}) and (\ref{eq:spec}).
The spectral density at zero momentum is shown in
Fig.~\ref{fig:kaon7} for several densities. As density increases the peak
 of the ${\bar K}$ pole moves towards lower
energies since the ${\bar K}$ optical potential becomes more attractive. 
The in medium $\Lambda(1405)$ resonance gets wider and, although not
signaled by any clear peak in the figure, its presence can be indirectly 
noticed from examining the ${\bar K}$ spectral density
on the right hand
side of the quasiparticle peak, which falls off more slowly than that on
the left hand side. 
We also notice some structure of the spectral
function to the left of the quasiparticle peak at energies
of the ${\bar K}$ of around $320-360$ MeV, the origin of which
can be traced back
to a pole of the free ${\bar K}N$ amplitude obtained with the J\"ulich
interaction in
the $L=1$, $I=1$ channel at an energy $\sqrt{s}=1230$ MeV, 
about 200 MeV
below the ${\bar K}N$ threshold. This pole comes from
the $\Sigma$ pole diagram of the bare J\"ulich potential which moves $35$
MeV up in
energy due to the iteration process involved in $T$-matrix. Hence,
the peak in the spectral
function is the in-medium reflection of this singularity in
the free space $T$-matrix.
The dotted line shows the spectral density at $\rho=\rho_0$ but keeping
only the $L=0$ component of the ${\bar K}N$ interaction. In agreement
with the behavior of the optical potential at zero momentum, we
observe
that the location of the quasiparticle peak only moves a few MeV, while
the width (height) gets reduced (increased) by about 30\%.

\section{Conclusions}
\label{sec:conclu}

We have performed a microscopic self-consistent calculation
of the  single-particle potential of a $\bar K$ meson embedded in 
symmetric nuclear matter, using the meson-exchange J\"ulich 
$\bar K N$ interaction.

Due to the strong energy dependence of the $\bar K N$ $G$-matrix it becomes
crucial to follow a self-consistent procedure to evaluate the
$\bar K$ self-energy. 
In this work we have analyzed two self-consistent schemes which produce
substantial different results. When only the real part of the ${\bar K}$
optical potential is retained in the self-consistent procedure, one
obtains
an attraction of about $-100$ MeV at zero momentum and a very small
imaginary part of around $-3$ MeV. This optical potential would lead to 
extremely narrow deeply bound kaonic states in nuclei. However, when the
complete complex optical potential is self-consistently determined, the
real part becomes 15\% less attractive and the imaginary part
increases to around $-25$ MeV, as a consequence of the widening of the
antikaon
strength which produces a ${\bar K}N$ amplitude smoother and  more spread
out over energies.

We have obtained the kaon optical potential as a function of the ${\bar
K}$ momentum, up to regions that are relevant
in the analysis of heavy-ion 
collisions where antikaons are created with a finite momentum.
Our results for the optical potential show a momentum dependence which is
moderate for the real
part and significantly more relevant for the imaginary part,
which is especially important for densities around the
saturation density.

We have also studied
the effect of including the higher partial waves of the ${\bar K}N$
interaction. Some influence is already seen at zero
${\bar K}$ momentum but the largest effects appear clearly at large
momenta, where the inclusion of the $L > 0$ partial waves of the
${\bar K}N$ interaction can
practically double the size of the optical potential. At a ${\bar 
K}$ momentum of 500 MeV/c the complex optical potential changes from the
$L=0$ 
value of
$(-28,-39)$ MeV to
$(-52,-61)$ MeV when all partial waves are included.

\section*{Acknowledgments}

We are very grateful to Monika Hoffmann for providing us with the
codes of the J\"ulich ${\bar K}N$ interaction and to E. Oset for useful
discussions.
This work is partially supported
by DGICYT contract number PB98-1247 and by the Generalitat de Catalunya
under project SGR98-11. One of us (L.T.) wishes to acknowledge support
from a doctoral fellowship of the Ministerio de Educaci\'on y Cultura (Spain).

\newpage
\renewcommand{\theequation}{\Alph{section}.\arabic{equation}}
\setcounter{section}{1}
\setcounter{equation}{0}
\section*{Appendix A}

In this appendix we show how to compute the angular average of
the Pauli operator, $Q_{\bar{K}N}$.
Defining $\vec{P}$ and $\vec{k}$ as the total and relative momenta of
the $\bar{K}N$ pair, respectively
\begin{equation}
\vec{P}=\vec{k}_{\bar{K}}+\vec{k}_{N} ,~~~~
\vec{k}=\frac{M_{N}\vec{k}_{\bar{K}}-M_{\bar{K}}\vec{k}_{N}}
{M_{\bar{K}}+M_{N}} \ ,
\end{equation}
we can rewrite the nucleon and antikaon momenta in the
laboratory system, $\vec{k}_{N}$ and $\vec{k}_{\bar{K}}$, as
\begin{equation}
\vec{k}_{N}=-\vec{k}+\frac{\xi}{1+\xi}\vec{P} , ~~~~
\vec{k}_{\bar{K}}=\vec{k}+\frac{1}{1+\xi}\vec{P} \ ,
\end{equation}
where a galilean transformation  has been 
used and $\xi=\displaystyle\frac{M_{N}}{M_{\bar{K}}}$.
The Pauli operator acts only on the nucleonic line and, in symmetric
nuclear matter, it reads
\begin{equation}
Q_{{\bar K}N}(\vec{P},\vec{k}) 
=\theta\left(\mid\frac{\xi}{1+\xi}\vec{P}-\vec{k}\mid-k_{F}\right) \ ,
\end{equation}
which depends on the angle between $\vec{P}$ and $\vec{k}$. In order to
eliminate this dependence,  we introduce
an
angle average
\begin{equation}
\overline{Q}_{{\bar K}N}(P,k)=\frac{1}{2} \int_{0}^{\pi} 
d\theta \sin{\theta}\,
Q(\vec{P},\vec{k}) \ ,
\end{equation}
such that the resulting function, given by
\begin{equation}
\overline{Q}_{{\bar K}N}(P,k)= \left\{ \begin{array}{cl}
\displaystyle\frac{1}{2}\left[1+\frac{k^2+\left(\frac{\xi}{1+\xi}\right)^2 
P^2-{k_{F}}^2}
{2kP\frac{\xi}{1+\xi}}\right]
& ~~{\rm for\ } \mid \frac{\xi}{1+\xi}P-k_{F}\mid < k <
\frac{\xi}{1+\xi}P+k_{F} , \\ 1 &
~~{\rm for\ }
k > \frac{\xi}{1+\xi}P+k_{F} \ {\rm or} \ k < \frac{\xi}{1+\xi}P-k_{F} ,
\\
0 & ~~\mbox{otherwise}
\end{array} \right. \ , \nonumber
\end{equation}
depends only on the modulus of the relative and total momenta of
the $\bar {K} N$ state.

\newpage
\renewcommand{\theequation}{\Alph{section}.\arabic{equation}}

\setcounter{section}{2}
\setcounter{equation}{0}
\section*{Appendix B}

In this appendix we show how to compute an appropiate average of
the ${\bar K}N$ center-of-mass momentum, $\vec{P}$, and the
nucleon momentum, $\vec{k}_{N}$, 
given an  external antikaon
momentum, $\vec{k_{\bar{K}}}$, and a  relative ${\bar K}N$  momentum,
$\vec{k}$, used as integration variable in Eq.~(\ref{eq:upot1}).
From Eqs.~(A.1) and (A.2) one obtains
\begin{equation}
\vec{P}=({1+\xi})(\vec{k}_{\bar K}-\vec{k}), ~~~
\vec{k}_{N}=\xi\vec{k}_{\bar K}-{1+\xi}\vec{k} \ .
\end{equation}
The angle average of the center-of-mass momentum is defined as
\begin{eqnarray}
  \overline {P^{2}}(k_{\bar{K}},k) &=& \frac{\displaystyle\int
d(\cos{\theta})\,
P^{2}(k_{\bar{K}},k,\cos{\theta})}{\displaystyle\int d(\cos{\theta})},
\end{eqnarray}
where
$P^{2}(k_{\bar{K}},k,\cos{\theta})=(1+\xi)^2(k_{\bar{K}}^{2}
+k^{2}-2k_{\bar{K}}k\cos{\theta})$, with $\theta$ being the angle between
$\vec{k}_{\bar{K}}$ and $\vec{k}$.
Similarly, for $\overline{k_{N}^{2}}$ we have
\begin{eqnarray}
  \overline {k^{2}_{N}}(k_{\bar{K}},k) &=& \frac{\displaystyle\int
d(\cos{\theta})\,
k^{2}_{N}(k_{\bar{K}},k,\cos{\theta})}{\displaystyle\int
d(\cos{\theta})},
\end{eqnarray}
where
$k^{2}_{N}(k_{\bar{K}},k,\cos{\theta})=\xi^2
k_{\bar{K}}^2+(1+\xi)^2k^2-2\xi(1+\xi)k_{\bar{K}}k \cos{\theta}$.

In both cases, the integration runs from $\cos{\theta_{m}}$ to 1,
where the expression for
$\cos{\theta_{m}}$ depends on two regions of integration, according to
the restrictions imposed by Pauli blocking, and is given by
\begin{equation}
\cos{\theta_{m}}= \left\{ \begin{array}{cl}
\displaystyle\frac{\xi^2 k_{\bar{K}}^2+(1+ \xi)^2k^2-k_{F}^2}
{2 \xi k_{\bar{K}}(1+ \xi)k}
& ~~{\rm for\ } \frac{|k_{F}- \xi k_{\bar{K}}|}{1+ \xi} \leq k \leq
\frac{k_{F}+ \xi k_{\bar{K}}}{1+ \xi}, \\ -1 &
~~{\rm for\ }
k \leq \frac{k_{F}- \xi k_{\bar{K}}}{1+ \xi}
\end{array} \right. \nonumber
\end{equation}

The resulting angle averages for $\vec{P}$ and $\vec{k}_{N}$ are given
by
\begin{equation}
\overline{P^{2}}(k_{\bar{K}},k)= \left\{ \begin{array}{cl}
(1+ \xi)^2 \left[ (k_{\bar{K}}^2+k^2)-
\displaystyle\frac{{[ \xi k_{\bar{K}}+(1+ \xi )k]}^2-k_{F}^2}
{2 \xi (1+ \xi )}\right]
& ~~{\rm for\ } \frac{|k_{F}- \xi k_{\bar{K}}|}{1+ \xi} \leq k \leq
\frac{k_{F}+ \xi k_{\bar{K}}}{1+ \xi}, \\
{(1+ \xi )}^2(k_{\bar{K}}^2+k^2) &
~~{\rm for\ }
k \leq \frac{k_{F}- \xi k_{\bar{K}}}{1+ \xi}
\end{array} \right. \nonumber
\end{equation}
\begin{equation}
\overline{k_{N}^{2}}(k_{\bar{K}},k)= \left\{ \begin{array}{cl}
 \xi^2  k_{\bar{K}}^2+ (1+ \xi ) ^2 k^2- 
\displaystyle\frac{{[ \xi k_{\bar{K}}+(1+ \xi )k]}^2-k_{F}^2}
{2}
& ~~{\rm for\ } \frac{|k_{F}- \xi k_{\bar{K}}|}{1+ \xi} \leq k \leq
\frac{k_{F}+ \xi k_{\bar{K}}}{1+ \xi}, \\
\xi^2 k_{\bar{K}}^2+{(1+ \xi )}^2k^2 &
~~{\rm for\ }
k \leq \frac{k_{F}- \xi k_{\bar{K}}}{1+ \xi}
\end{array} \right. \nonumber
\end{equation}

\begin{figure}[htb]
\centerline{
     \includegraphics[width=0.6\textwidth]{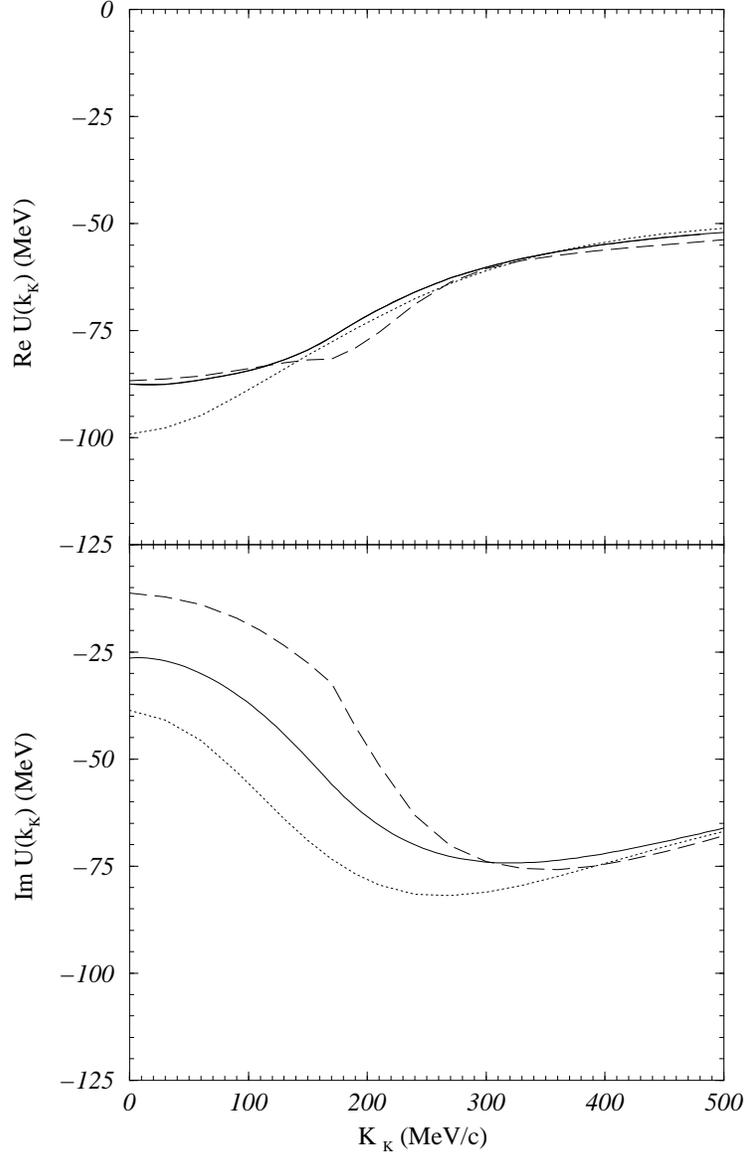}
}
      \caption{\small
Real and imaginary parts of the ${\bar K}$ optical potential
as functions of the antikaon momentum for various approximations. Solid
line: our results. Dashed line: nucleon recoil corrections neglected.
Dotted line: nucleon recoil corrections neglected and free--particle
spectrum used for nucleons.}
        \label{fig:kaon6}
\end{figure}

\end{document}